\documentclass{Interspeech2024}

\interspeechcameraready

\title{Mmm whatcha say? Uncovering distal and proximal context effects in first and second-language word perception using psychophysical reverse correlation}

\name[affiliation={1,2}]{Paige}{Tuttösí}
\name[affiliation={3}]{H. Henny}{Yeung}
\name[affiliation={3}]{Yue}{Wang}
\name[affiliation={3}]{Fenqi}{Wang}
\name[]{Guillaume} {Denis}
\name[affiliation={2}]{Jean-Julien}{Aucouturier}
\name[affiliation={1}]{Angelica}{Lim}

\address{
  School of Computing Science$^{1}$, Dept. of  Linguistics$^{3}$, Simon Fraser University, Canada \\
  Université de Franche-Comté, SUPMICROTECH, CNRS, Institut FEMTO-ST$^{2}$, France
}
\email{ptuttosi@sfu.ca}


\begin{document}

\maketitle

% the abstract here must exactly match the abstract entered into the paper submission system
\begin{abstract}
Acoustic context effects, where surrounding changes in pitch, rate or timbre influence the perception of a sound, are well documented in speech perception, but how they interact with language background remains unclear. Using a reverse-correlation approach, we systematically varied the pitch and speech rate in phrases around different pairs of vowels for second language (L2) speakers of English (\textipa{/i/}-\textipa{/I/}) and French (\textipa{/u/}-\textipa{/y/}), thus reconstructing, in a data-driven manner, the prosodic profiles that bias their perception. Testing English and French speakers (\textit{n}=25), we showed that vowel perception is in fact influenced by conflicting effects from the surrounding pitch and speech rate: a congruent proximal effect 0.2s pre-target and a distal contrastive effect up to 1s before; and found that L1 and L2 speakers exhibited strikingly similar prosodic profiles in perception. We provide a novel method to investigate acoustic context effects across stimuli, timescales, and acoustic domain.% and opens the use of reverse-correlation prosodic profiles to improve the comprehension for those with nonnative/reduced comprehension ability. 
\end{abstract}

\section{Introduction}
 In human-to-human interaction, the ability of a speaker to adapt to an interlocutor is invaluable. Humans will modify their speech to interlocutors with reduced comprehension abilities, e.g., babies \cite{McClay22} or second language (L2) speakers \cite{Redmon20}. Moreover, taking cues from multiple modalities, a speaker is able to perceive when they are not being understood and make adjustments to their speech production to increase clarity, both in terms of linguistic \cite{Biro22, Maniwa09} and paralinguistic (e.g. emotional \cite{Banse96, Bachorowski99}) contexts. While synthesized speech is quickly approaching human speech clarity and naturalness, it often still lacks the ability to adapt to interlocutors, especially in a controllable manner. A progression towards adaptive synthesised speech is to first understand, in a fine grained manner, how speech can be adjusted to facilitate perception.

One classical view on vowel perception is that it is a local process categorizing speech sounds by comparison with a pre-learned auditory representation, presumably a spectral one in formant space \cite{LIB00}. Speech sounds, however, are highly variable within and across speakers and it is  widely documented that word perception also operates relative to its acoustic context \cite{Stilp20, MCMU11}. For instance, following a phrase that is spoken quickly, a sound can be perceived to be longer than it actually is \cite{Newman96}.

There is debate, however, about the exact temporal characteristics of such context effects in ecological sentences, and whether the locus of contextual influences depends on the speech cue that is involved \cite{Reinisch13}. For instance, \cite{Newman96} suggested that the influence of preceding speech rate on phoneme distinction may be limited to a temporal window of one or two adjacent phonemes, while \cite{Johnson99} suggest that spectral context effects result from a form of speaker’s vocal tract length normalization, which benefits from exposure to long-term spectral cues accumulated over possibly several sentences. Whether and how such proximal and distal effects can interact or compete, and the exact timescale at which they occur, remain poorly understood \cite{Stilp20}. 

Determining the acoustic and temporal characteristics of information intake in sentence perception is methodologically challenging for speech-perception research. While hypothesis-driven experimental paradigms can establish the causal influence of specific cues on word contrasts by systematically varying their intensity (e.g. 13 distal speech rates on the perception of a target word in \cite{Heff13}), assessing the relative weight of several temporal regions of interest becomes quickly impractical \cite{Newman96}. To document such temporal dynamics, several studies have relied on eye-tracking in visual search tasks (e.g. printed words, or a `visual world' paradigm, where objects corresponding to each word are shown), and compared the time course of eye fixations to the occurrence of contextual cues \cite{SHATZ06,Reinisch13}. However, this type of study can be underpowered, and there may be other methodologies that permit the automatic extraction, in a data-driven manner, of a listener’s mental representation of what acoustic profile (i.e. what cues, and where) drives a specific speech sound contrast in one direction or another. 

In this work, we investigate the use of a classic experimental method, psychophysical \textbf{reverse-correlation} \cite{AHU71}, which has seen a recent surge of interest in speech perception research for its ability to uncover an individual's mental representation of what prosodic pattern drives, e.g., judgements of a speaker’s dominance \cite{Ponsot18} or confidence \cite{Goupil21}. Specifically, we use a phase-vocoder technique to systematically and concurrently vary the pitch and speech rate in phrases surrounding pairs of vowels/words that are known to be difficult for L2 speakers: English (\textipa{/i/}-\textipa{/I/}) and French (\textipa{/u/}-\textipa{/y/}); we then use reverse correlation to reconstruct the prosodic profiles that bias the perception of these word pairs in one direction or the other. 

To illustrate the ability of this procedure to identify fine differences between groups of participants, we collected data from bilingual French and English speakers, both on the same French and English stimuli. The research literature is equivocal on whether non-native language processing is able to recruit similar acoustic context mechanisms as in native language. For instance, Kang, Johnson, and Finley \cite{KANG16} found that French-L2 speakers failed to use vowel context (i.e. to compensate for coarticulation) when judging fricative sounds when such vowels were unfamiliar; but others have found differing results \cite{VIS13}. Here, we specifically ask what surrounding pitch and speech rate cues L1 speakers use to differentiate pairs of sounds, and whether L2 speakers are sensitive to the same prosodic profiles.

%We compare these profiles for L1 and L2 speakers of English and French.  %Accordingly, we ask: \textbf{how can the perception of vowels that are known to be difficult for L2 speakers}, both in terms of perception and production, \textbf{be controlled and perception be improved thorough the aforementioned modifications of pitch and duration across a word or phrase?}

\section{Procedure \& Stimulus generation}

\subsection{Procedure}

Reverse correlation is an experimental paradigm aimed at discovering the signal features that govern a participant's judgement by analysing their responses to large sets of stimuli whose acoustic characteristics have been systematically manipulated \cite{ADOL16}. Specifically, we presented participants with a series of 250 trials, each consisting of a single base recording containing an ambiguous target word. For each trial, the base recording was manipulated with a different random profile of pitch and speech rate, and participants were asked which of two target words they heard (1-interval, 2-alternative forced choice). Responses are then analysed to reconstruct the prosodic profile that maximizes the likelihood of responding to one option or the other (for a review, see \cite{Mur11}). 

This study included 4 different experiments: (A) two involving random manipulations of the isolated target word (English or French), aimed at establishing a baseline for the intrinsic pitch and rate of the two alternatives, and (B) two involving manipulations of  phrases containing the target word, aimed at uncovering extrinsic acoustic context effects. When participants took part in more than one experiment, the order was randomized across language and type of stimuli (word and phrase). For each experiment, the order of response options (e.g. ``pill/peel" or ``peel/pill") was randomized across participants. Each experiment lasted, on average, 12 minutes. 

\subsection{Participants} 

N=114 participants took part in the study: N=54 French-L1 speakers (female: 24, M=32.4yo ± 9.7) and N=60 English-L1 speakers (female: 33, M=30.0yo ± 10.7). Each experiment (2 word tasks, 2 phrase tasks) was carried out with \emph{n}=25 French-L1 and
\emph{n}=25 English-L1 speakers, some of whom participated in more than one experiment. Participants had
a self-rated proficiency in the L2 language (English or French) ranging from 2-5 (1: no proficiency, 5: fluent) with a mode of 4. English participants were
recruited primarily in anglophone Canada from Simon Fraser University and French participants in France from SUPMICROTECH-ENSMM, as well as on Prolific. The study received internal ethics approval from both universities.

\subsection{Target words}

In each language, we selected pairs of vowels that are known to be difficult for English or French L2 speakers. For English, we selected \textipa{/i/} and \textipa{/I/}. The vowel \textipa{/i/}, as in ``beat" \textipa{/bit/}, is a high, tense vowel that exists in both French and English. In contrast, vowel \textipa{/I/} , as in ``bit" \textipa{/bIt/}, is lower and lax, and does not exist within the French phonetic inventory. Often, French-L1 speakers learning English will replace \textipa{/I/} by \textipa{/i/}, for example, saying 
``sheep" \textipa{[Sip]} when they mean to say ``ship" \textipa{/SIp/} \cite{iverson12}. 

For French, we selected the vowels \textipa{/u/} and \textipa{/y/}. Once again, this pair of vowels is known to be difficult for English-L1-French-L2 speakers in both perception and production \cite{Flege87, Levy09}. The vowel \textipa{/u/}, such as in ``fou" \textipa{/fu/} (mad), is a high, back vowel, and exists in both French and English. In contrast, the vowel \textipa{/y/}, such as in ``fût" \textipa{/fy/} (cask), is a high, front vowel, and does not exist within the English lexicon. English speakers tend to overcompensate for the lack of \textipa{/y/} vowel in their language production and often replace \textipa{/u/} with \textipa{/y/}, such as mispronouncing ``beaucoup" \textipa{/boku/} (a lot) with the rather immodest phrase \textipa{[boky]} \cite{STU13}.

We selected words that had the same beginning and ending consonants across languages to maintain consistency in confounding factors such as co-articulation. The selected words were ``pill" \textipa{/pIl/} and ``peel" \textipa{/pil/} in English, and ``pull" \textipa{/pyl/} and ``poule" \textipa{/pul/} in French.

\subsection{Phrase stimuli}

For our phrase experiments, we used the phrase ``I heard them say" (FR: ``je l'ai entendu dire") preceding the word in an attempt to control contextual bias. The phrases were generated using the Hugging Face interface for CoquiXTTS\footnote{https://huggingface.co/spaces/coqui/xtts}. The language was set to English for the English phrase and French for the French phrase. An L1 male reference voice was provided and no other modifications were made to the TTS settings.

\begin{figure}[t]
  \centering
  \includegraphics[width=180pt]{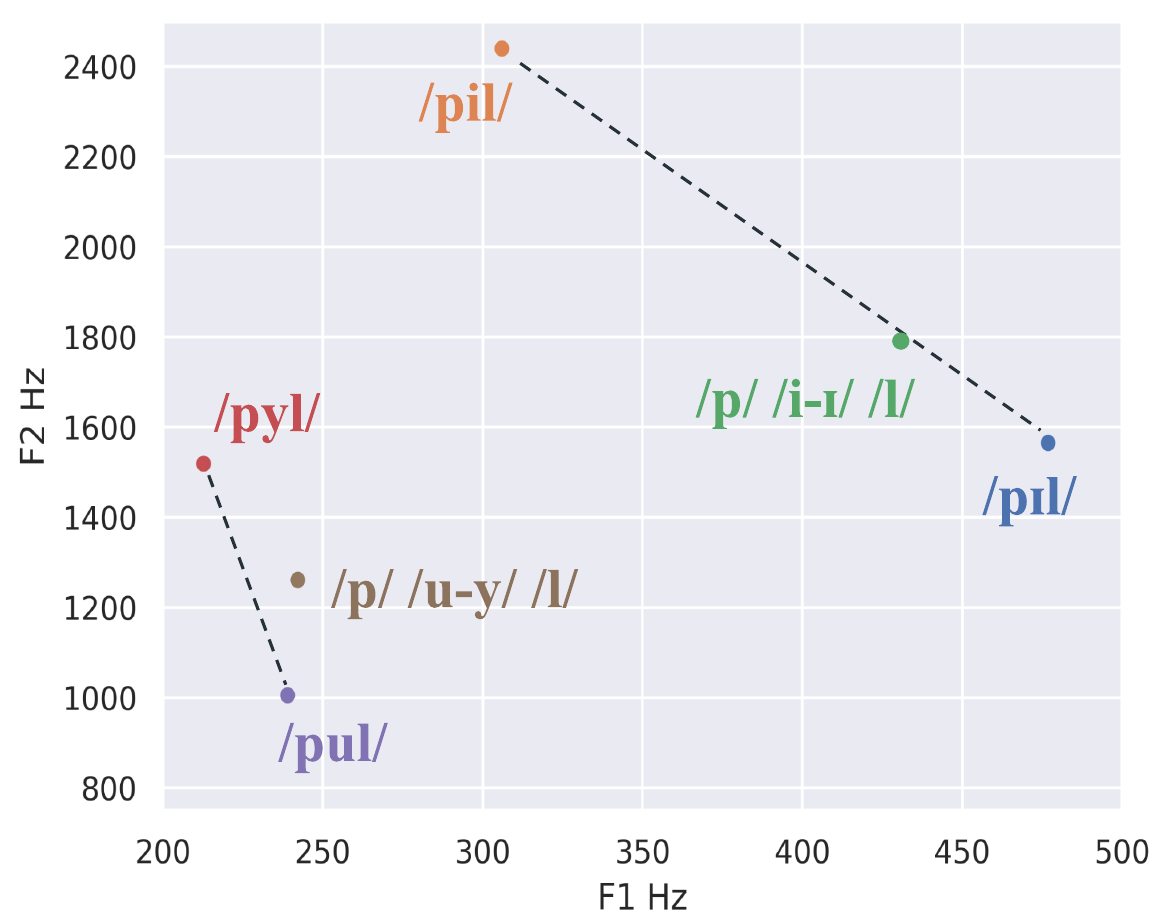}
  \caption{F1-F2 plot of initial and final formants for ambiguous vowels /u-y/ and /i-I/.}
  \label{fig:formants}
\vspace*{-5mm}
\end{figure}

\subsection{Stimulus manipulation}

To control for response bias in the 1-interval, 2-alternative task (i.e. one alternative being judged more common than the other), we generated morphed sounds intermediate between each of the two vowel pairs \textipa{/i/}-\textipa{/I/} and \textipa{/u/}-\textipa{/y/}. To do so, we used the Praat software \cite{praat} to modify the F1 and F2 formants of each original target word, increasing their resemblance by steps of 10Hz (original formants: \textipa{/i/} F1: 305.89Hz, F2: 2440.77Hz; \textipa{/I/} F1: 476.85Hz, F2: 1565.45Hz; \textipa{/u/} F1: 238.90Hz, F2: 1005.67Hz; \textipa{/y/} F1: 212.61Hz, F2: 1519.49Hz). We then modified the Whisper ASR algorithm \cite{radford2022robust} (v20231117, medium multilingual) to access the log-probability of the two possible word interpretation at every step in the formant grid; using the difference in these log-probabilities to select the transformation step that resulted in the smallest difference in prediction probability between the two target words. The resulting formants can be seen in Fig.~\ref{fig:formants} with the final \textipa{/i/}-\textipa{/I/} vowel having F1: 436.77Hz, F2: 1722.68Hz (synthesized from ``peel" \textipa{/pil/}), and the final \textipa{/u/}-\textipa{/y/} vowel having F1: 238.13Hz, F2: 1258.68Hz (synthesized from ``poule" \textipa{/pul/}). These ambiguous words served as base stimuli in the word-task, and were inserted manually in their original phrase (at a zero-crossing 120ms after the end of the last word ``say/dire'') to serve as base stimuli for the phrase-task.

Second, we generated the reverse-correlation stimuli from these ambiguous base sounds using the open-source CLEESE toolbox \cite{BURR19}. A voice transformation toolbox that creates random fluctuations around an audio file’s original contour of pitch and speech rate. The pitch contour of the recordings was artificially flattened to a constant 120Hz. We then transformed the stimuli by randomly manipulating their pitch and duration independently in \emph{n} successive windows of 100ms (\emph{n}=4 for words; \emph{n}=13 for phrases), each of which with a factor sampled from a normal distribution (pitch: $\mu$=0, $\sigma$=100 cents (i.e. 1 semitone); duration: $\mu$=0\%$,\sigma$=1\% (i.e. doubling or halving the window's duration); both distributions clipped at $\pm 2 \sigma$). These values were chosen so as to cover the range observed in naturally produced utterances and were linearly interpolated between successive time points to ensure a natural sounding transformation.

\section{Results}

\subsection{Validation of ambiguity}
\label{bias}
We explored our success at creating ambiguous sounds using our combined PRAAT/Whisper approach by analysing participant response bias. Because random manipulations were centered on zero, we expected a 50\% response rate for all alternative options. For the word experiments, English-L1 speakers answered ``peel" $52\%$ of the time and ``poule" $64\%$ of the time; French-L1 speakers answered ``peel" $54\%$ of the time and ``poule" $70\%$ of the time. For phrases, English speakers responded with $59\%$ ``peel" and $59\%$ ``poule", while French-L1 speakers' responses were $54\%$ ``peel" and $66.6\%$ ``poule". It appears that it was more difficult to manipulate the original sound for native French speakers in French, especially for a lone word; discussions with participants concurred with these results.

\begin{figure}[t]
  \centering
  \includegraphics[width=\linewidth]{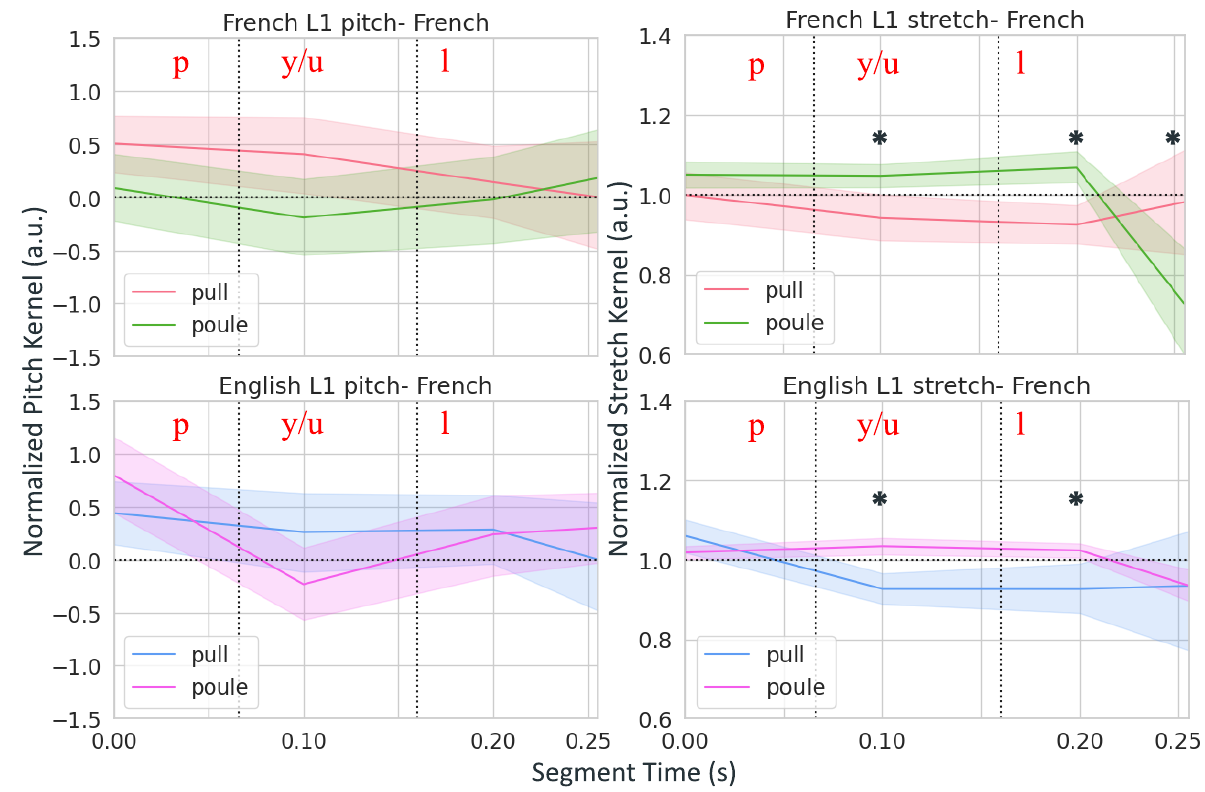}
  \caption{Pitch (left) and speech rate (right) kernels for French-L1 (top) and L2 (bottom) speakers for the French words ``pull" and ``poule". In all figures, colored areas mark 95\% confidence intervals on the mean, and * marks time segments that differ statistically at $\alpha=0.05$. Smaller values for the rate kernel mean shorter duration, i.e. {\it faster} speech rate. }
  \label{fig:french-word}
  \vspace*{-5mm}
\end{figure}

%Although we do see the origin word being selected more often, the overall class balance is not particularly affected. The largest imbalance is seen in the French word for French-L1 speakers with a $70-30\%$ class balance, which, in most cases, will not affect reverse correlation experiments and did not lead to extreme variance in our results.

\subsection{Word reverse-correlation}

For each participant and each word task (EN/FR), we computed first-order kernels from reverse-correlation data using the {\it classification image} method \cite{Mur11}. We computed the average random pitch and speech rate transformation profile of the recordings classified as one response option (e.g. ``pill''), and compared it with the average pitch and speech rate profile of the recordings classified as the other response (e.g. ``peel''). Kernels were normalized by dividing them by the root-mean-square sum of their values. For each participant and each task, this procedure resulted in two dim=4 vectors of pitch and rate of speech values for words, representing what pitch and speech rate transformations should be applied to a given base word in order to increase the likelihood of recognizing a particular target. In the following, we test for differences between pitch and speech rate kernel values in each task using paired t-tests at every time point.  

\emph{French words}. Our prediction was that French-L1 (FL1) speakers' perception of \textipa{/y/}, a high/front vowel, would be driven by higher pitch and faster speech rate compared to \textipa{/u/}. Our results confirmed this (Fig.\ref{fig:french-word}) for speech rate (FL1 - 0.1s: t(24)=2.80, p=.010; 0.2s: t(24)=3.52, p=.002), and pitch (at 0s and 0.1s, albeit non-statistically). French-L2 (EL1) shared the same pattern of data, with a non-statistical pitch increase at t=0.1s and faster speech rate at 0.1s: t(24)=4.57, p$<$.001; and 0.2s: t(24)=2.73, p=.012. Although statistically weak, this pattern of results (higher pitch/faster rate for \textipa{/y/}) was confirmed by segments located on the target word in the phrase kernels.

\begin{figure}[t]
  \centering
  \includegraphics[width=\linewidth]{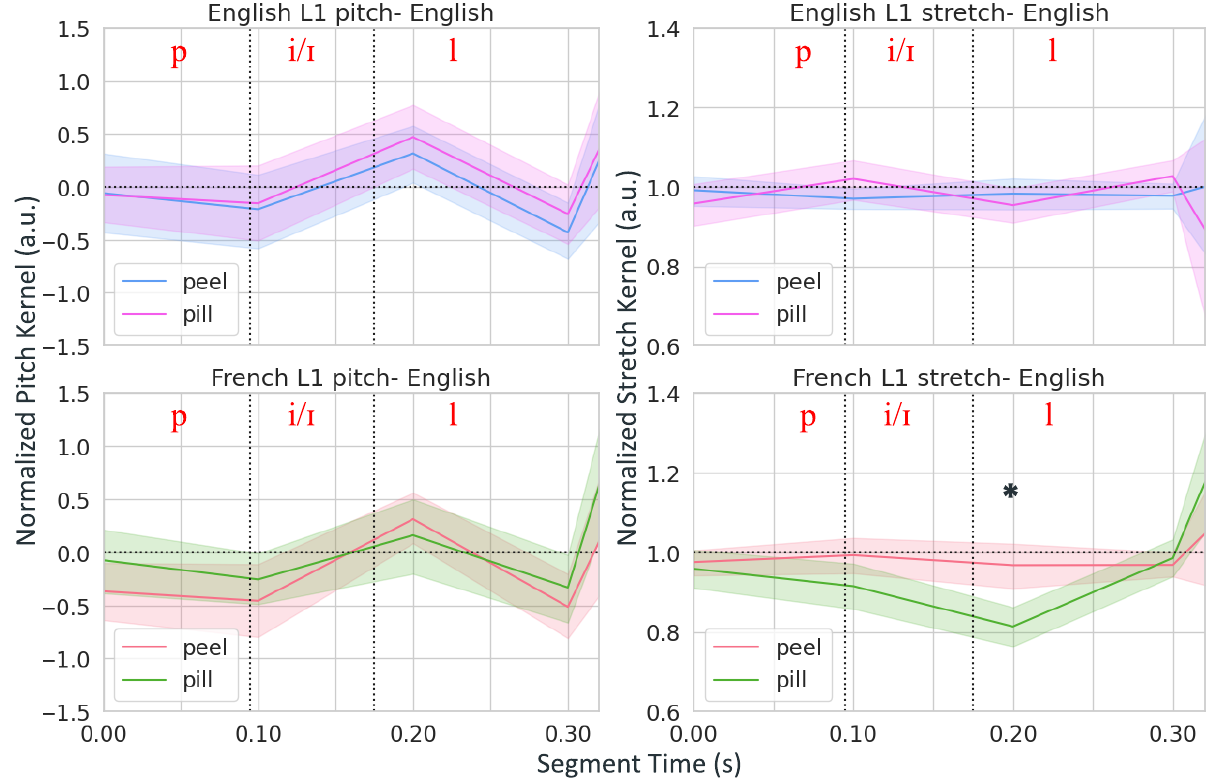}
  \caption{Pitch (left) and speech rate (right) kernels for English-L1 (top) and L2 (bottom) speakers for the English words ``peel" and ``pill". Smaller values for the rate kernel mean shorter duration, i.e. {\it faster} speech rate.}
  \label{fig:english-word}
  \vspace*{-5mm}
\end{figure}

\emph{English words}. Our prediction was that EL1 speakers' perception of \textipa{/i/}, a high/tense vowel, would be driven by higher pitch and slower speech rate compared to \textipa{/I/}. Our results did not confirm this prediction (Fig.\ref{fig:english-word}): 
while the speech rate kernels for \textipa{/i/} showed a significantly slower rate for L2 speakers (0.2s: t(24)=3.13, p=.005); for pitch, the kernels were not significant and, if anything, indicated the opposite, a preference for increased pitch in \textipa{/I/} for L1 speakers. Again, this pattern of results (higher pitch and a slower rate for \textipa{/i/}) was confirmed by segments located on the target word in the phrase kernels.  

\subsection{Phrase reverse-correlation}

For each participant and phrase task (EN/FR), we computed reverse-correlation kernels using the same procedure as for the words, resulting in dim=13 kernels for pitch and speech rate.

\begin{figure}[t]
  \centering
  \includegraphics[width=\linewidth]{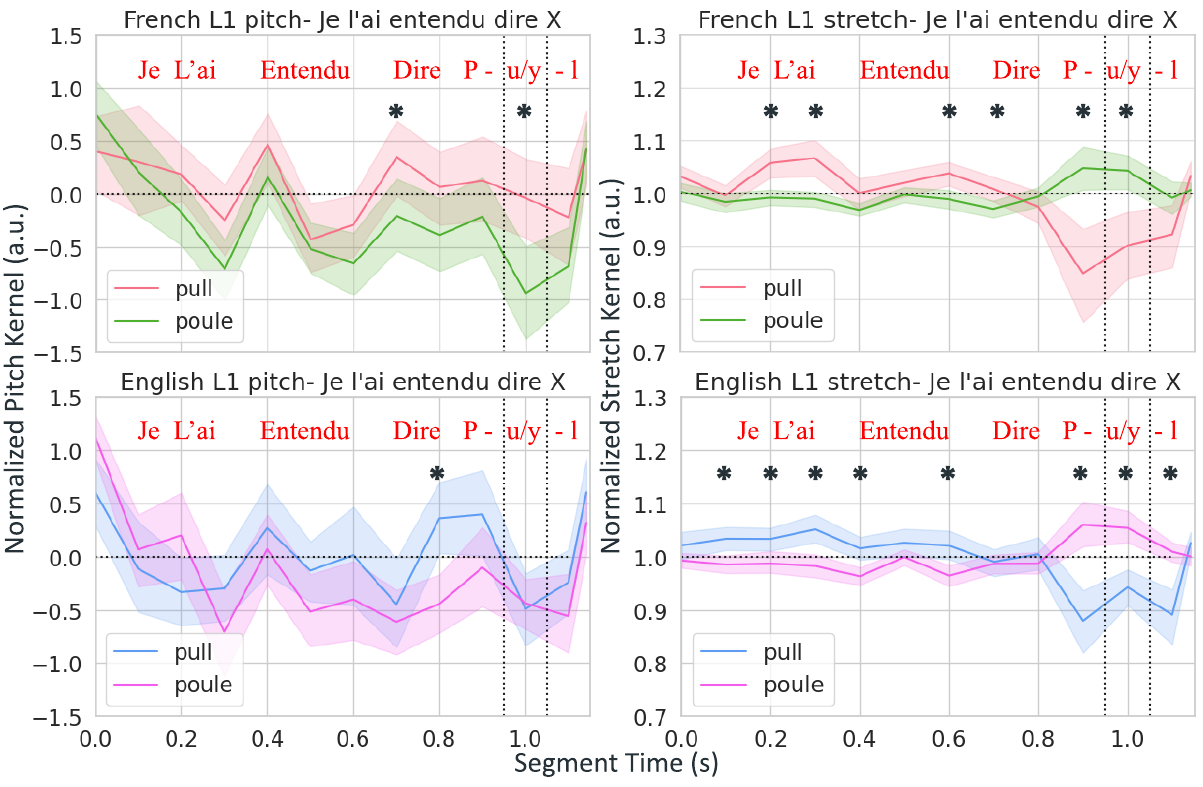}
  \caption{Pitch (left) and speech rate (right) kernels for French-L1 (top) and L2 (bottom) speakers for the French phrases containing ``pull" and ``poule". Smaller values for the rate kernel mean shorter duration, i.e. {\it faster} speech rate.}
  \label{fig:french-phrase}
  \vspace*{-5mm}
\end{figure}

\emph{French phrases}. Phrase reverse-correlation data confirmed, this time significantly, that within the target word hearing \textipa{/y/} is driven by higher pitch in FL1 (1s: t(24)=-2.63, p=.015) and faster speech rate in both FL1 and EL1 speakers (FL1 - 1.0s: t(24)=2.99, p=.006; EL1 - 1.0s: t(24)=4.04, p$<$.001, 1.1s: t(24)=3.71, p=.001). Regarding the surrounding context of the word, the literature predicts the existence of contextual contrast effects, i.e. that \textipa{/y/} is driven by {\it lower} pitch and speech rate before the target word. In actuality, our data revealed a mix of contrastive and congruent contextual influences: pitch (Fig. \ref{fig:french-phrase}-left) was not associated with any contrastive effect, but rather a proximal congruent effect 200-300ms pre-target (FL1- 0.7s: t(24)=-2.18, p=.040; EL1- 0.8s: t(24)=-4.05, p$<$.001), i.e. an increase of pitch immediately before the target word biases the response towards the higher-pitch alternative \textipa{/y/}. Speech rate (Fig. \ref{fig:french-phrase}-right) exhibited the expected long-term (distal) contrastive effect (FL1 - 0.2s: t(24)=-3.72, p=.001, 0.3s: t(24)=-3.86, p=.001, 0.6s: t(24)=-2.89, p=.008, 0.7s: t(24)=-2.24, p=.018; EL1 - 0.1s: t(24)=-3.35, p=.003, 0.2s: t(24)=-2.53, p=.018, 0.3s: t(24)=-4.12, p$<$.001, 0.4s: t(24)=-2.99, p=.006, 0.6s: t(24)=-3.46, p=.002), where a slower speech rate at the beginning of a phrase and a proximal congruent effect 100ms pre-target (FL1 - 0.9s: t(24)=3.25, p=.003; EL1 - 0.9s: t(24)=3.76, p=.001) biases the response towards the faster alternative \textipa{/y/}, resulting in an overall scissor-shape profile. This pattern of result was remarkably conserved in L2 speakers (Fig. \ref{fig:french-phrase}-bottom).    

\emph{English phrases}. As above, the effect of pitch and rate of speech manipulation within the target word was consistent with the word task: contrary to theoretical predictions, \textipa{/I/} is driven by higher pitch, significantly in L1 speakers (0.9s: t(25)=-2.69, p=.013) and, in L2 speakers, by the expected faster speech rate (0.9s: t(24)=2.15, p=.042, 1.0s: t(24)=3.61, p=.001). Contrary to the French results, pitch showed contrastive effects for L1 speakers within the phrase, both distally (0.1s: t(25)=3.23, p=.003, 0.3s: t(25)=3.68, p=.001) and in the immediate proximity of the target word (0.8s: t(25)=2.91, p=.007). Speech rate for both L1 and L2 speakers showed the same scissor-shape pattern as the FR phrases, with distal contrastive effects (EL1 - 0.1s: t(25)=-2.52, p=.018, 0.2s: t(25)=-2.31, p=.030, 0.4s: t(25)=-3.47, p=.002, 0.5s: t(25)=-2.34, p=.028; FL1 - 0.1s: t(24)=-3.02,p=.006, 0.3s: t(24)=-3.53,p=.002, 0.4s: t(24)=-3.07, p=.005, 0.5s: t(24)=-2.75, p=.011), and a strong proximal congruent effect (EL1: 0.7s: t(25)=3.77, p=.001, 0.8s: t(25)=5.98, p$<$.001, however, the rate difference on the target word was not significant for EL1; FL1: 0.8s: t(24)=4.25, p$<$.001). As before, this pattern of result was remarkably conserved across L1 (Fig. \ref{fig:french-phrase}-top) and L2 speakers (Fig. \ref{fig:french-phrase}-bottom). 
%We do not see this proximal congruent effect in the first language speakers, however, this is not due to the speech rate at the proximal segment (0.7s: t(25)=3.77, p=.001, 0.8s: t(25)=5.98, p$<$.001), which moves in the expected direction, but rather to the unexpected reversal from the expected speech rate within the word. 

\begin{figure}[t]
\vspace*{-2mm}
  \centering
  \includegraphics[width=\linewidth]{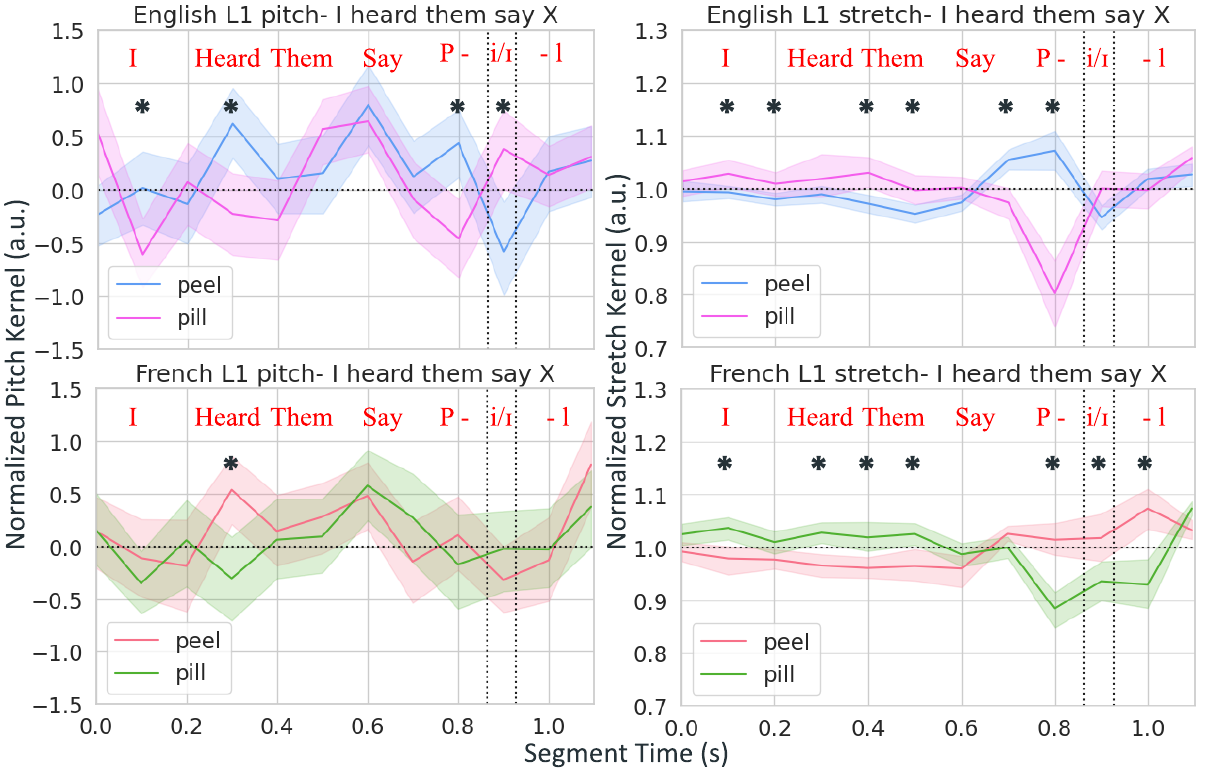}
  \caption{Pitch (left) and speech rate (right) kernels for English-L1 (top) and L2 (bottom) speakers for the English phrases containing ``peel" and ``pill". Smaller for the rate kernel mean shorter duration, i.e. {\it faster} speech rate.}
  \label{fig:english-phrase}
  \vspace*{-5mm}
\end{figure}

\section{Discussion \& Limitations}
This study provides a proof of concept for using the reverse-correlation paradigm to uncover basic phonetic findings, such as the effects of intrinsic pitch or surrounding speech rate on vowel recognition. In four experiments with N=114 French and English participants, we found systematic acoustic context effects across 2 word pairs and 2 sentences that were, in general, consistent with predictions from the speech perception literature, notably for distal contrast effects \cite{Stilp20}. 

Compared to traditional hypothesis-driven experimental paradigms, reverse correlation offers several key advantages. First, it allows for uncovering the precise chronometry of context effects and aligning cues with specific phrase elements. In particular, we found that contrastive context effects became congruent proximal to the target, with a tipping point around 200-300ms pre-target. Second, while speech sound perception typically integrates multiple cues and is context dependent \cite{MCMU11}, phonetic studies are often concentrated on examining a single cue at a time. Reverse correlation offers a methodology to explore combinations of cues at once in a single experiment, producing results more akin to every day human perception. Future work could include additional audiovisual cues, such as formant manipulation \cite{PONS18} or orofacial gestures \cite{LIU22}, as well as quantifying the relative perceptual weight of each cue, taking inspiration from feature selection approaches \cite{GARG19}.

Despite this potential, the work remains preliminary and suffers from several limitations. First, we saw weaker and less-consistent effects for pitch than speech rate, especially within the FR phrase and in the FL1 group. Since French is not a language with lexically specified word stress, it is possible that pitch cues are less important for FL1 than EL1 speakers. Another possible explanation may be that the selected phrase ``je l'ai entendu dire", with e.g., more consonants than the English phrase, lends itself comparitvely poorly to pitch transformations. Future work should therefore investigate the generalizability of the results with more word pairs and more phrase contexts. Second, results in the EN word task (preference
for increased pitch in /I/ for L1 speakers) were inconsistent with theoretical predictions of higher intrinsic pitch for \textipa{/pil/}. While this effect was replicated in the phrase task, and was consistent with the contrast effects obtained in the rest of the phrase, it would be interesting to clarify why this occurred and examine possible individual differences in how participants combine pitch and spectral cues in this sound. For example, it is possible that EL1 speakers weigh spectral cues higher than speech rate or pitch cues, whereas the L2 speakers rely more heavily on prosody to disambiguate vowel tensity  \cite{YLI10,RED20}. Finally, our experiments used a 1-interval, 2-alternative design, in an attempt to keep the experiment short and feasible in an online setting. However, such as design potentially introduces response bias (section \ref{bias}) and other decision variables that can obfuscate purely phonetic mechanisms (see e.g. pitch kernels that significantly depart from zero in the same direction for {\it both} response options, Figure \ref{fig:english-phrase}-left), and it would be interesting to reproduce these results with a longer, 2-interval 1-alternative task.

\section{Acknowledgements}
This work was supported by NSERC Discovery Grant 06908-2019, the France Canada Research Fund, the Mitacs Globalink Research Award, and the Fondation Pour l’Audition (FPA RD-2021-12). The authors thank P.~Maublanc, R.~Guha, and A.~Adl Zarrabi for their valuable discussions; V.~Yang, B.~Burkanova, C.~Zhang, and M.~Durana for their help running our study; and the Rajan Family for their support. This work has been conducted in the framework of the EIPHI Graduate school (ANR-17-EURE-0002 contract).

\bibliographystyle{IEEEtran}
\bibliography{mybib}

\end{document}